\documentclass[12pt]{iopart} 

	\usepackage{graphicx}
\begin{document}

	\title[Importance of interlinguistic similarity and ...]{Importance of interlinguistic similarity and stable bilingualism when two languages compete}

	\author{J Mira$^1$, L F Seoane$^1$ and J J Nieto$^2$}

	\address{$^1$ Departamento de F\'isica Aplicada, Universidade de Santiago de Compostela, 15782 Santiago de Compostela, Spain.}
	\address{$^2$ Departamento de An\'alise Aplicada and Instituto de Matem\'aticas, Universidade de Santiago de Compostela, 15782 Santiago de Compostela, Spain. }
	\ead{jorge.mira@usc.es}

\begin{abstract}

	In order to analyze the dynamics of two languages in competition, one approach is to fit historical data on their numbers of speakers with a mathematical model in which the parameters are interpreted as the similarity between those languages and their relative status. Within this approach, we show here, on the basis of a detailed analysis and extensive calculations, the outcomes that can emerge for given values of these parameters. Contrary to previous results, it is possible that in the long term both languages coexist and survive. This happens only when there is a stable bilingual group, and this is possible only if the competing languages are sufficiently similar, in which case its occurrence is favoured by both similarity and status symmetry.

\end{abstract}

\pacs{89.65.-s}
\maketitle

	An aspect of globalization that alarms many is the replacement of local tongues by more hegemonic languages \cite{Sutherland}, a trend that has been investigated from multiple points of view, including that of Physics \cite{Hawking-Gell-Mann, Schulze-Stauffer, Castello-Eguiluz, Crystal, Abrams-Strogatz, Mira-Paredes}. Where two languages compete, it is reasonable to assume that outcome is influenced by their relative status \cite{Abrams-Strogatz} (i.e. the perception by the speaker of the social and/or economic advantages each language offers) and their similarity \cite{Mira-Paredes} (or the complementary concept, interlinguistic distance \cite{Nowak-Komarova}), but consideration of these factors has been hindered by difficulties in their operationalization. 

	One of the earliest and simplest mathematical models of language shift was that of Abrams and Strogatz \cite{Abrams-Strogatz}, who considered a stable population in which two languages with different statuses compete for speakers. This model, which involves the analysis of the evolution of the number of speakers along time, predicted that one of the languages would inevitably die out and was successfully fitted to historical data on competition between Scottish Gaelic and English, Welsh and English, and Quechua and Spanish, among other language pairs \cite{Abrams-Strogatz}. However, it did not take into account the possibility of bilingual individuals, a possibility that is of course realized in numerous multilingual societies. In Spain, for example, where Castilian Spanish is the official language throughout the State but in certain regions is co-official with another language (mainly Galician, Basque, Catalan or Valencian), individual bilingualism is common in communities with more than one co-official language. 

\begin{figure}
			\includegraphics[width=\textwidth]{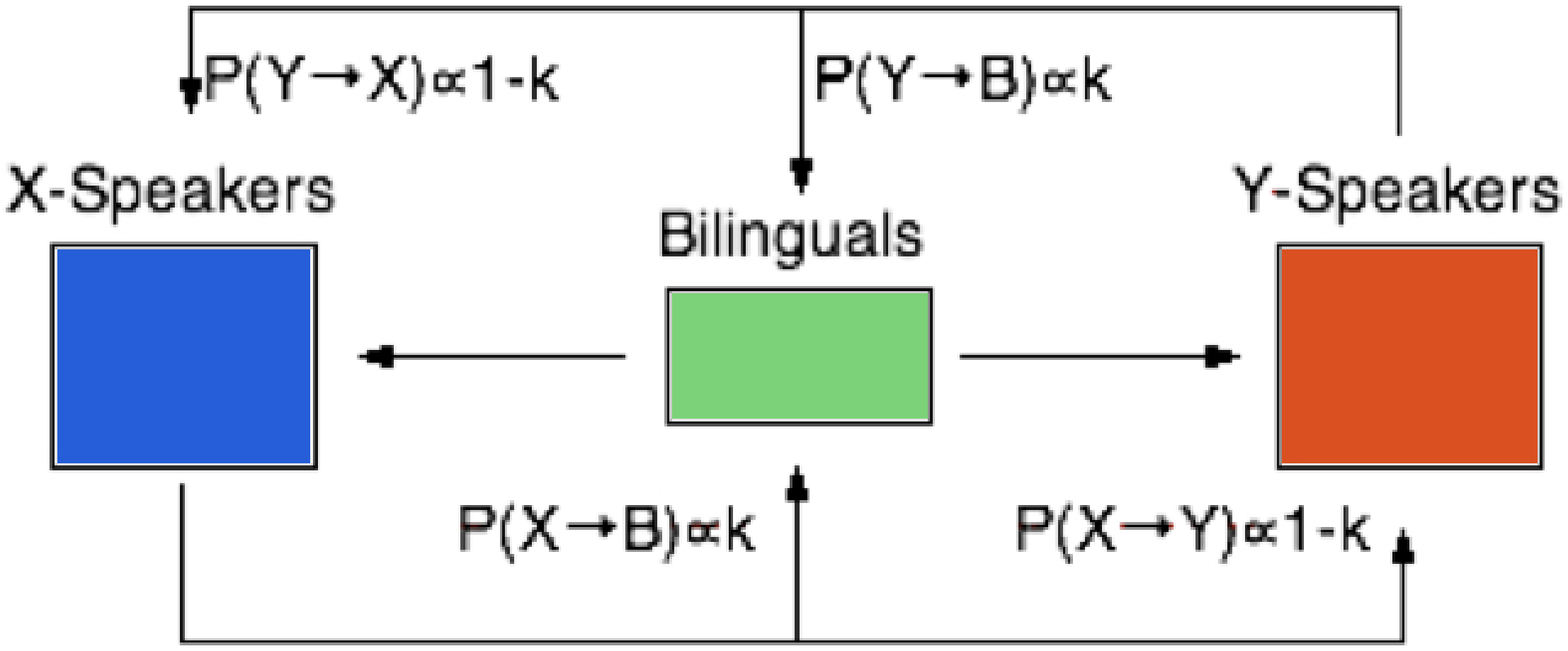} 
			\caption{\textbf{Flux among X-monolingual, Y-monolingual and bilingual groups. } According to the model set forth in equations (\ref{eq:1})-(\ref{eq:3}) and (\ref{eq:4})-(\ref{eq:7}), flux is governed both by the relative statuses of the competing languages (the relative social and/or economic advantages they offer) and by interlinguistic similarity $k$, the probability that the disappearance of a monolingual speaker of one language will be compensated by the appearance of a bilingual rather than by a monolingual speaker of the other language. }
			\label{fig:0}
		\end{figure}

	We recently showed that the historical evolution of the use of Galician and Castilian in Galicia (N.W. Spain) can be explained by a modified Abrams-Strogatz model that allows for bilingual as well as monolingual speakers of the competing languages, and which includes a parameter that represents the ease of bilingualism \cite{Mira-Paredes} (figure \ref{fig:0}). We considered a population in which monolingual X-speakers make up a fraction $x$, monolingual Y-speakers a fraction $y$, and bilinguals (B) a fraction $b$ (with $xÊ+ÊyÊ+ÊbÊ=Ê1$), and the dynamics of language change are accordingly described by the system
		\numparts
		\begin{equation}
			{\rmd x \over \rmd t} = yP_{YX} + bP_{BX} - x(P_{XY} + P_{XB}), 
			\label{eq:1}
		\end{equation}
		\begin{equation}
			{\rmd y \over \rmd t} = xP_{XY} + bP_{BY} - y(P_{YX} + P_{YB}), 
			\label{eq:2}
		\end{equation}
		\begin{equation}
			{\rmd b \over \rmd t} = xP_{XB} + yP_{YB} - b(P_{BY} + P_{BX}); 
			\label{eq:3}
		\end{equation}
		\endnumparts
where $P_{XY}$ denotes the probability of a monolingual X-speaker being replaced in the population by a monolingual Y-speaker, with analogous notation for the other possible replacements. The probability of a monolingual being replaced by a mono- or bilingual speaker of the other language is assumed to be proportional both to the status of the second language, i.e. the social and/or economic advantages it offers, and to a power of the proportion of the population that speak it. Thus, denoting by $s$ the relative status of language X and by $1-s$ that of language Y,
		\numparts
		\begin{equation}
			P_{XB} = c\cdot k(1-s)(1-x)^a, 
			\label{eq:4}
		\end{equation}
		\begin{equation}
			P_{YB} = c\cdot ks(1-y)^a, 
			\label{eq:5}
		\end{equation}
		\begin{equation}
			P_{BX} = P_{YX} = c\cdot (1-k)s(1-y)^a 
			\label{eq:6}
		\end{equation}
		\begin{equation}
			P_{BY} = P_{XY} = c\cdot (1-k)(1-s)(1-x)^a; 
			\label{eq:7}
		\end{equation}
		\endnumparts
where $c$ is a normalization factor related to the time scale, $a$ is the power parameter, and $k$ is the probability that the disappearance of a monolingual speaker of X (respectively Y) will be compensated by the appearance of a bilingual rather than by a monolingual speaker of Y (respectively X). We identify interlinguistic similarity with this parameter $k$. 

	Note that since people do not actually forget their native tongue(s), the above model is a model of population renewal, rather than of individuals switching from one linguistic practice to another, except that it also allows monolinguals to become "born-again" bilinguals. When $kÊ=Ê0$ it reduces to the Abrams-Strogatz model \cite{Abrams-Strogatz}, or decays towards this model if $b$ is initially nonzero. Note also that population growth does not invalidate the model, so long as the various linguistic groups are all affected in proportion to their size. 
	
\begin{figure}
			\includegraphics[width=\textwidth]{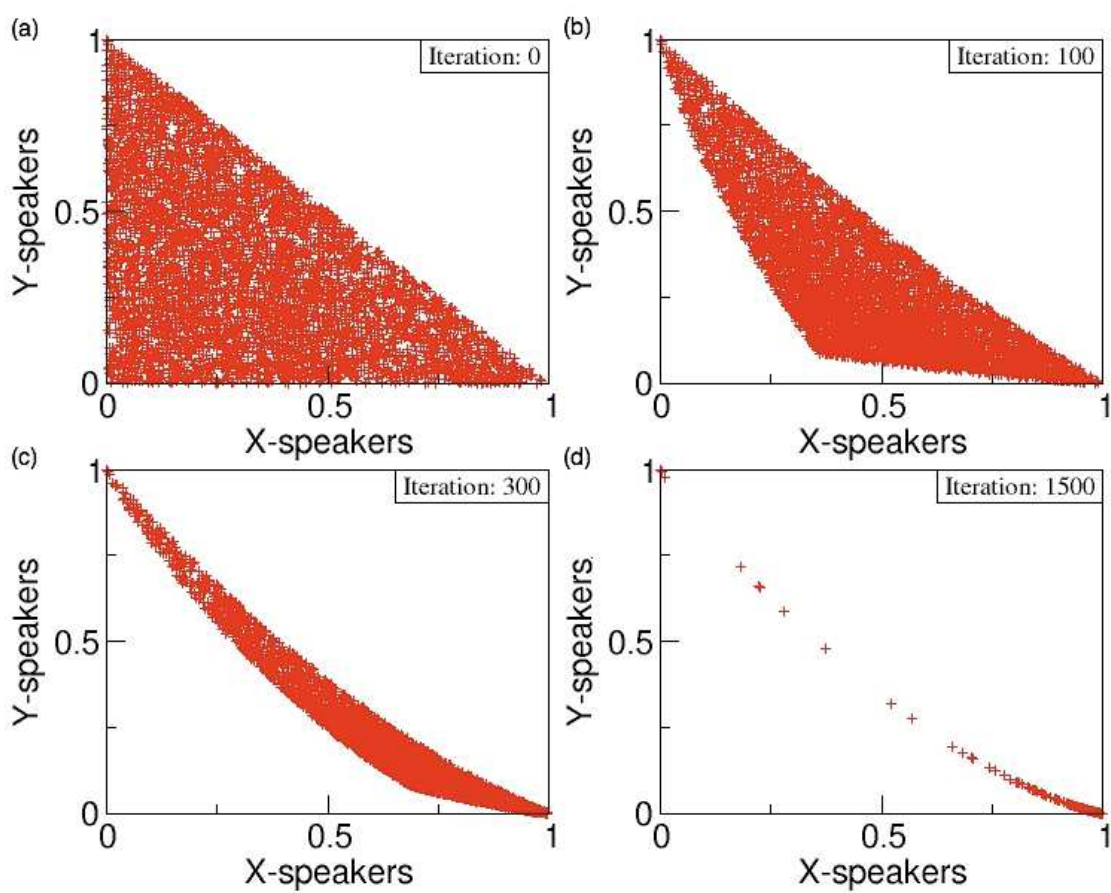} 
			\caption{\textbf{Evolution of language dominance when languages X and Y compete.} Example with $sÊ=Ê0.75$ and $kÊ=Ê0.3$. The graph axes represent the proportions of monolingual speakers of X and Y in the population. \textbf{a}, Random distribution of initial states. \textbf{b-d}, State distributions attained after respectively $100$, $300$ and $1\>500$ steps of the computation. }
			\label{fig:1}
		\end{figure}

\begin{figure}
			\includegraphics[width=\textwidth]{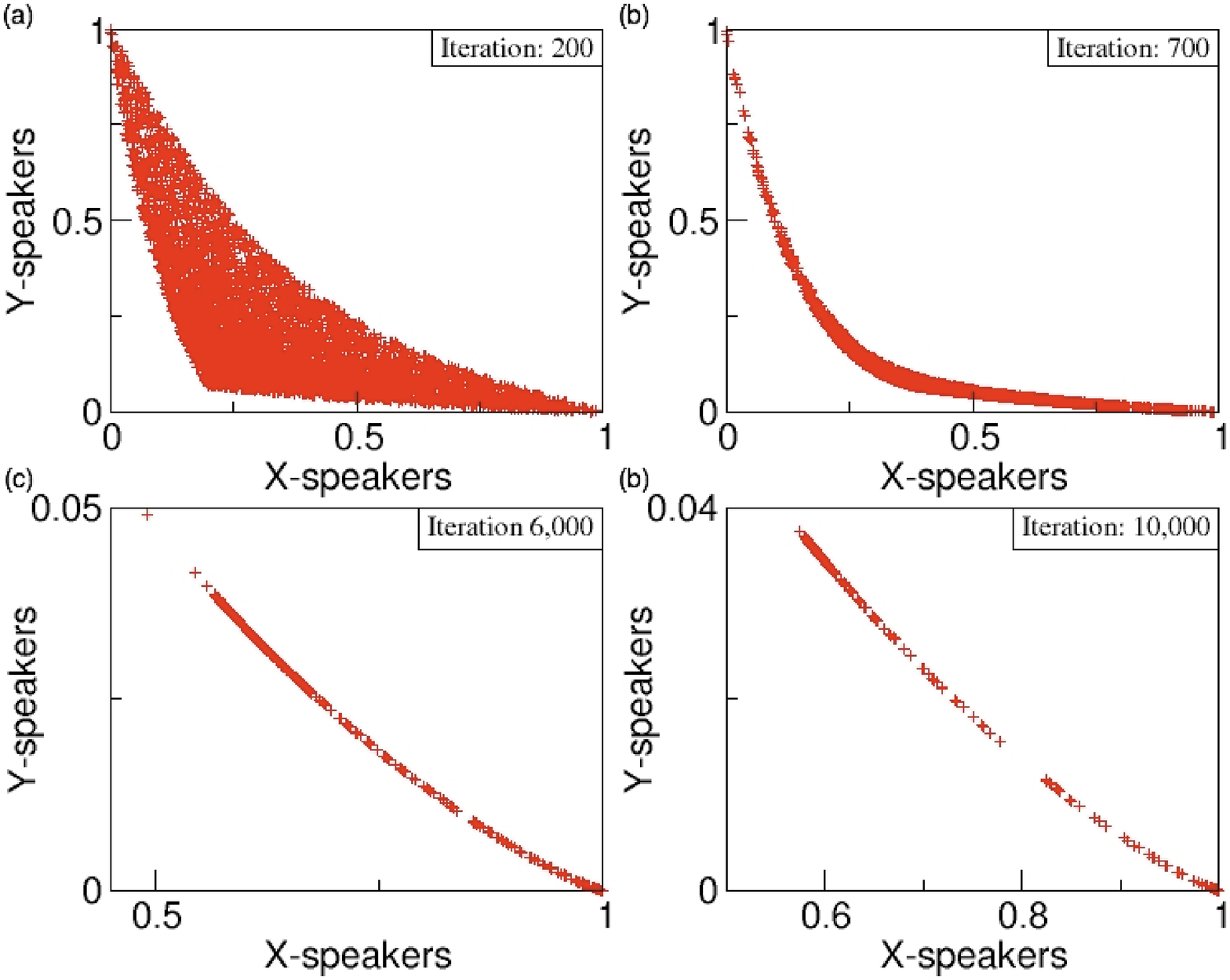} 
			\caption{\textbf{A case with two different attractors.} State distributions attained after  \textbf{a}, $200$; \textbf{b}, $700$; \textbf{c}, $6\>000$; and \textbf{d}, $10\>000$ computation steps when $sÊ=Ê0.67$ and $kÊ=Ê0.77$ (the small regions shown in panels c and d contain all the points of the distributions). Depending on the initial proportions of X-monolinguals, Y-monolinguals and bilinguals, language Y dies out completely or survives, in the latter case mostly among bilinguals. Note the opening of a gap in the points of panel d, because the states evolve towards different attractors $(x=1, y=0)$ and $(x= 0.6, y= 0.04)$. }
			\label{fig:2}
		\end{figure}

	Although the model sketched above adequately accounts for the Galician data up to the present \cite{Mira-Paredes}, the question arises whether this situation is stable. More generally, what are the possible long-term outcomes of competition between two languages, and under what conditions do they come about? Namely, might two languages coexist stably? To investigate these issues we have carried out extensive calculations, systematically varying $s$ and $k$, to determine population states $(x,y)$ that act as point attractors for the coupled system defined by equations (\ref{eq:1})-(\ref{eq:3}) and (\ref{eq:4})-(\ref{eq:7}) with $xÊ+ÊyÊ+ÊbÊ=Ê1$; the problem is well-posed because, as it is easily shown, the velocity field on the boundary of the set of possible states (defined by $xÊ\geÊ0$, $yÊ\geÊ0$, $xÊ+ÊyÊ\leÊ1$) never takes the system outside this set, whatever the values of $s$ and $k$. 
	K
	Initially, our calculations were performed as follows. For all $(s,k)$ of the form $(0.05n_s, 0.05n_k)$ $(0Ê\leÊn_s,n_kÊ\leÊ20)$, $10\>000$ starting states $(x_0,y_0)$ were randomly selected in the set of possible states, and the discretized form of the above system (\ref{eq:1})-(\ref{eq:3}) was solved numerically for each, with $aÊ=Ê1.31$, using a time step $\Delta tÊ=Ê0.01$ (the value $aÊ=Ê1.31$ was chosen because Abrams and Strogatz \cite{Abrams-Strogatz} found that this parameter was surprisingly constant, $1.31Ê\pmÊ0.25$, when their model was fitted to $42$ data sets concerning dissimilar language pairs). The $10\>000$ calculations proceeded concurrently, and were halted as soon as all the state points had converged to within a circle of radius $10^{-5}$. Figures  \ref{fig:1}(a)-(d) show selected stages of this process for the case ($sÊ=Ê0.75$, $kÊ=Ê0.3$). However, it was soon found that there were values of $(s,k)$ for which the state points did not converge, and examination of their distribution indicated that this was due to the existence of more than one point attractor, the attractor tended to by any particular state point depending on its starting state. In these cases the calculations were allowed to proceed for times several orders of magnitude longer than the average single-attractor convergence time, until all the state points were within two or three circles of radius $10^{-5}$ (figureÊ \ref{fig:2} shows the result for the case $sÊ=Ê0.67$, $kÊ=Ê0.77$).

\begin{figure} 
			\includegraphics[width=\textwidth]{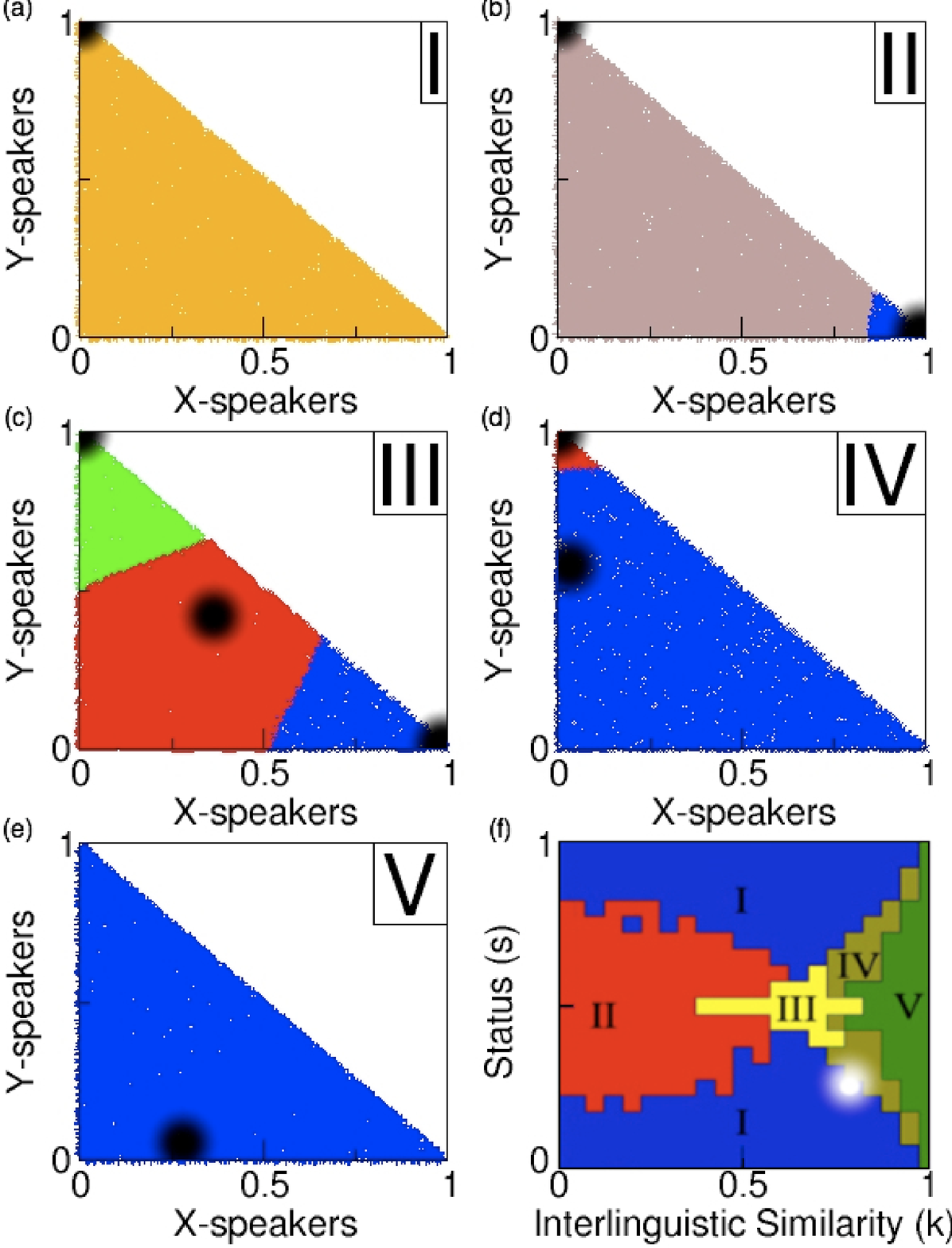} 
			\caption{\textbf{The five possible stable situations determined by $s$ and $k$.} See text for descriptions of types I-V. Random distributions of initial states for typical cases of \textbf{a}, type I ($sÊ=Ê0.80$, $kÊ=Ê0.65$); \textbf{b}, type II ($sÊ=Ê0.40$, $kÊ=Ê0.20$); \textbf{c}, type III ($sÊ=Ê0.50$, $kÊ=Ê0.65$); \textbf{d}, type IV ($sÊ=Ê0.35$, $kÊ=Ê0.75$); and \textbf{e}, type V ($sÊ=Ê0.65$, $kÊ=Ê0.85$). Attractors are represented as large black spots. Where multiple attractors exist, initial states evolving towards different attractors are shown in different colours. \textbf{f}, The areas of the $s,k$-plane occupied by systems of types I-V. The point $k,s$ of the system Galician-Castilian is marked with a white spot. }
			\label{fig:3}
		\end{figure}

\begin{figure} 
			\includegraphics[width=0.5\textwidth]{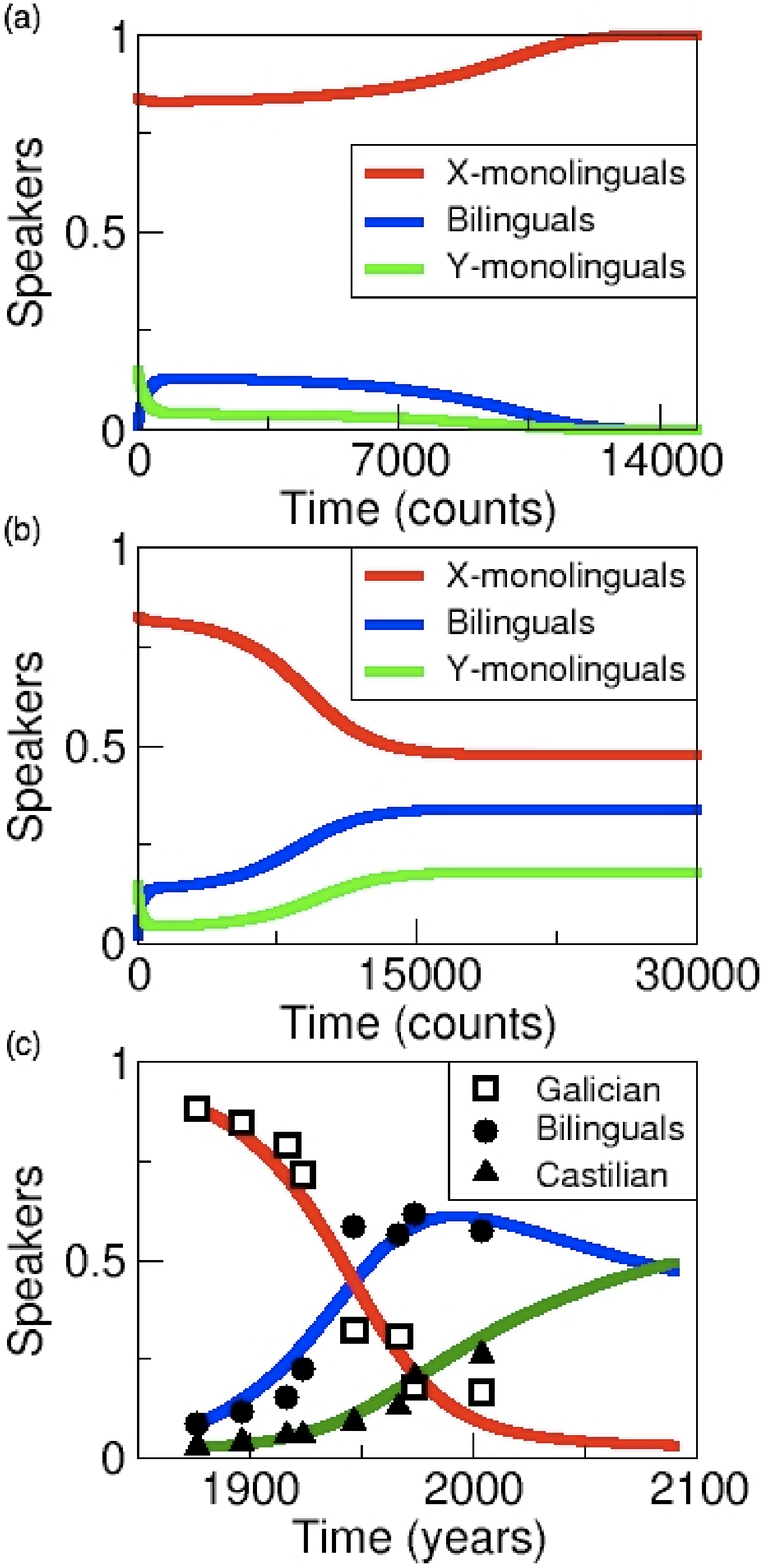} 
			\caption{\textbf{Time evolution of single cases.} Time dependence of the proportions of X-monolinguals (red), Y-monolinguals (green) and bilinguals (blue). Panels \textbf{a} and \textbf{b} illustrate for a type II system ($sÊ=Ê0.53$, $kÊ=Ê0.55$) how the fate of the system can depend critically on initial conditions, language Y becoming extinct if initially $xÊ= 0.84$ and $yÊ=Ê0.15$ (a) but not if $xÊ= 0.83$ and $yÊ=Ê0.15$ (b). \textbf{c}, Results of fitting the model to historical data for Galician (X, red), Castilian (Y, green) and bilinguals (B, blue) in Galicia (N.W. Spain). Although the fitted values of $s$ and $k$, $0.26$ and $0.80$, place Galicia in zone I of figure \ref{fig:3}(f), thus predicting the eventual extinction of Galician, Galician-Castilian bilinguals are not expected to disappear within this century or the next. }
			\label{fig:4}
		\end{figure}

	Depending on $s$ and $k$, the following five situations were observed to emerge: 
		\begin{itemize} 
			\item I) There is just a single stable state at $xÊ=Ê1$ or $yÊ=Ê1$, i.e. one of the languages becomes extinct even though it may initially have been dominant. This is the behaviour illustrated in figuresÊ \ref{fig:1}(a)-(d) and in figure \ref{fig:3}(a). 
			\item II) There are stable states at both $xÊ=Ê1$ and $yÊ=Ê1$; one of the languages dies out, but which one depends on the initial distribution $(x_0,y_0)$, as illustrated in figure \ref{fig:3}(b). 
			\item III) There are stable states at $xÊ=Ê1$ and $yÊ=Ê1$, together with a third that lies below the line $xÊ+ÊyÊ=Ê1$ and thus corresponds to the presence of a stable bilingual group (figure \ref{fig:3}(c)). 
			\item IV) There is a stable state lying below the line $xÊ+ÊyÊ=Ê1$ (i.e. with a non-empty bilingual group) together with just one stable monolingual state at $xÊ=Ê1$ (figures \ref{fig:2}(a)-(d)) or $yÊ=Ê1$ (figure \ref{fig:3}(d)). 
			\item V) There is just one stable state, and it includes a stable bilingual group (figure \ref{fig:3}(e)). 
		\end{itemize} 
Note that in cases II, III and IV there are sharp boundaries between the initial state zones leading to different outcomes; at these boundaries, an exogenous injection of just a few speakers into one group or another can determine whether a language lives or dies, as is illustrated in figures \ref{fig:4}(a) and \ref{fig:4}(b). 

	Figure \ref{fig:3}f shows which of the above five situations was led to by each $(s,k)$ value. The salient aspects of this map are that the stable existence of a bilingual group requires that $k$ exceed a minimum value of about $0.35$; that if $kÊ\geÊ0.6$, then the less symmetric languages X and Y are statuswise, the larger $k$ must be for stable bilingualism; and that when the two languages are moderately symmetric statuswise, which of them disappears depends on the initial sizes of the linguistic groups if they are essentially dissimilar ($kÊ\le0.4$), but not if they are more similar ($0.4Ê<ÊkÊ<Ê0.6$). 

	Returning to the case of Galician and Castilian in Galicia, the inclusion of recently published data \cite{RAG} in the analysis (figure \ref{fig:4}(c)) corroborates our previous estimates \cite{Mira-Paredes} of $s_{Galician}$ ($0.26$) and $k$ ($0.80$), in fairly good agreement with lexicostatistics-based estimates of similarity among closely related Romance languages \cite{Dyen-Kruskal}. These values place Galicia in zone I of figure \ref{fig:3}(f), and accordingly predict the eventual extinction of Galician. However, the close proximity of zone IV suggests, and figure \ref{fig:4}(c) shows, that extinction is not imminent: in fact, it is predicted that by the end of the century the population will be roughly equally divided between Castilian monolinguals and bilinguals.

	The model used in this work has two evident limitations. Firstly, the model does not consider possible alterations of the relative proportions of the linguistic groups due to immigration, emigration, or differential birth and/or death rates. Secondly, partially related to the previous one, is that the relative status of the two languages may well vary in time. Also, of course, it remains to be seen whether the definition of linguistic similarity in terms of the dynamics of the population renewal process corresponds to any purely linguistic concept; the assumption that similarity is symmetric, i.e. that $P_{XB}/[(1-s)(1-x)^a]Ê=ÊP_{YB}/[s(1-y)^a]$, is clearly an idealization; and there is the issue of whether two highly similar languages really have separate identities, especially when they have similar status and one or both only survive among \emph{bilingual} speakers. The present results nevertheless suggest that the competition between two languages does not inevitably lead to the extinction of one of them. \\

\ack 
	The authors wish to acknowledge the help of Dr. \'Angel Paredes from the Institute for Theoretical Physics, University of Utrecht.

\end{document}